\documentstyle[12pt]{article}
\newcommand{\NP}[1]{Nucl. \ Phys.}
\newcommand{\PL}[1]{Phys. \ Lett.}
\newcommand{\p}[1]{\partial}
\newcommand{\PRL}[1]{Phys.\ Rev.\ Lett. }
\newcommand{\AP}[1]{Ann.\ Phys. }
\newcommand{\IM}[1]{Inv.\ Math. }
\newcommand{\JMP}[1]{ J.\ Math.\ Phys. }
\newcommand{\RMP}[1]{Rev.\ Mod.\ Phys.}
\newcommand{\JDG}[1]{ J.\ Diff.\ Geom. }
\newcommand{\PTP}[1]{ Prog.\ Theor.\ Phys. }
\newcommand{\SPTP}[1]{Suppl.\ Prog.\ Theor.\ Phys. }
\newcommand{\PR}[1]{Phys.\ Rev. }
\newcommand{\PREP }[1]{Phys.\ Reports }
\newcommand{\NC}[1]{Nuovo \ Cim. }
\newcommand{\NCL }[1]{Nuovo\ Cim.\ Lett. }
\newcommand{\CMP}[1]{Commun.\ Math.\ Phys. }
\newcommand{\TMF}[1]{Theor.\ Math.\ Phys. }
\newcommand{\CQG}[1]{Class.\ Quant.\ Grav. }
\newcommand{\MPL}[1] { Mod. Phys. Lett. }
\newcommand{\IJMP}[1] { Int. J. Mod. Phys. }
\newcommand{\RMAP}[1] { Rep.\ Math.\ Phys. }
\begin{document}

\title{Non-critical NSR string field theory\\ and \\ discrete states
interaction
in 2D supergravity.}
\author{ I.Ya.Aref'eva \thanks{E-mail: AREFEVA@QFT.MIAN.SU}
\\and \\ A.P.Zubarev \thanks{Supported in part by Moscow Physical Society
Grant}\\ Steklov Mathematical Institute,
Vavilov 42, GSP-1,117966, Moscow  }
\date {February, 1991}

\maketitle
\begin{abstract}

String field theory  for the non-critical NSR string is described.
In particular it gives string field theory
for the 2D super-gravity coupled to a $\hat{c}=1$ matter
field. For this purpose double-step pictures changing operators for
the non-critical NSR string are
constructed. Analogues of the critical supersymmetry transformations are
written for $D<10$,
they form a closed on-shell algebra, however their action on vertices is
defined only for discrete value of the Liouville momentum. For D=2 this means
that  spinor massless field  has its superpartner
in the NS sector only if its  momentum is fixed.

Starting from string field theory
we calculate string amplitudes. These amplitudes for D=2 have poles which are
related with discrete set of primary fields, namely 2R$\to$2R
amplitude has poles corresponding to the n-level NS excitations with discrete
momenta $p_1=n,~~p_2=-1\pm (n+1)$.
\end{abstract}
\vspace{5.0
cm}
$$~~~~~~~~~~~~~~~~~~~~~~~~~~~~~~~~~~~~~~~~~~~~~~~~~~~~{\bf SMI-1-92}~$$

\newpage
\section   {Introduction}
The goal of this paper is to present a string field theory associated with
the continuum super-Liouville theory. We will use the free field representation
\cite {GN,DHK} and  the modified Witten string field theory
\cite {WSFT,AMZ,PTY}.

String field theory is supposed to be background
independent and, therefore, to give a framework for discussing
nonperturbative effects.
Considerable progress in understanding non-perturbative
effects was achieved for 2-dimensional gravity coupled
to a $c=1$ matter \cite{M}-\cite{Yang} within the frameworks of
two  approaches to this problem: matrix models \cite{M} and
the continuum Liouville theory  \cite {Pol,GrKl,Polch}.
It is thus of interest to construct a
string field theory in the context of these models as well
as  their super generalizations. For bosonic case
the string field theory associated with matrix models was constructed in
 \cite {DasJ} and with the continuum Liouville theory in  \cite {AZ}.
 String field theory  for $D<26$ non-critical bosonic string was presented in
 \cite {PT}.

 2D string theories are characterized by the
 existence of a ring of discrete primary
 fields  \cite {WGR}. Discrete states have found in matrix model calculations
\cite {Gross,DM} and continuum calculations of tachyon amplitudes
\cite {Pol,AZ,DiFr,Yang} and are related with non-trivial cohomology of the
BRST
charge \cite {Lian,Pilch,IO}.

One of the outstanding nonperturbative problems of string theory
is that of supersymmetry breaking. It is therefore of great interest to
formulate lower dimensional model of superstrings. Supermatrix
models are discussed in \cite{SMM}.

 We investigate string states in both NS sector and  R sector in the context
 of string field theory. According to rough estimations, in the NSR string
 after the GSO projection
there is only one degree of freedom which corresponds to  massless spinor
field.
However a host of extra discrete states appears in the careful analysis of the
quantization procedure in the light-cone gauge.
This analysis shows that in the NS sector there are only discrete states
and in the R sector there are massless spinor field and discrete states.
So one gets an asymmetrical spectrum (the massless spinor and 2D vector field
cannot
form a supermultiplet) and an operator of supersymmetry does not exist.
Nevertheless analogues of the critical supersymmetry transformations can
be written for $D<10$,
they form a closed on-shell algebra, however their action on vertices are
defined only for discrete value of the Liouville momentum. In this sense spinor
state with fixed
value of momentum $|R;~ p_{1}=0~,p_{2}=-1>$ has its superpartner in the bosonic
sector,
it is the discrete state $|NS;~p_{1}=0,~p_{2}=0>$. In the bosonic sector we
have tachyon field
(massless field after the redefinition of momentum) however this field
 should be excluded after the GSO projection.
It would be interesting to find the corresponding
discrete set of states in the matrix model in the super case.

String field theory provides a
systematic method for calculation of states interaction. In particular,
the string field action gives an action for the massless spinor field, which is
a
fermionic analogue of the Das-Jevicki-Polchinski (DJP) action. One cannot say
that this action is a superpartner of the  DJP action since the DJP action
describes the
tachyon action and the tachyon  should be excluded after the GSO projection.

String field theory gives also a regular method for calculation of amplitudes.
We construct 4-points amplitudes in the different sectors.
RR$\to$RR amplitude has poles
which correspond to the on-shell NS discrete states with momentum-energy
$(p_1,p_2)$ equal to $p_1=n, ~p_2 =-1 \pm (n+1),~$ ($ n=0,1,2,...~$).

The paper is organized as follows. First, we review some well-known fact
about non-critical NSR string. Then we present an explicit formula for
picture-changing operators. These operator will allow us to examine string
states in  different pictures.
For completeness we briefly present the light-cone analysis of spectrum and
in more details we
investigate questions connected with supersymmetry.
 We then present string field action and discuss how fermionic analogue
 of the Das-Jevicki-Polchinski action can be extracted from it.
 We end with the derivation of general formulae for
N-point on-shell spinor amplitudes and the detail discussion of  properties
of the 4-point on-shell amplitudes.

\section{Non-critical NSR string.}
\subsection{Notations}
In the free field representation \cite{GN,DHK}
the first quantized non-critical NSR string action in the conformal gauge
has the form
\begin{equation}
S=-\frac{1}{8\pi}\int d^2\zeta \sqrt{\hat{g}}({\hat{g}}^{\alpha \beta}
\partial _{\alpha }X_{\mu}\partial_{\beta}X_{\mu}~+~i\hat{R}Q_{\mu}X_{\mu}~-~
\frac{i}{2}\psi_{\mu}\gamma^{\alpha}D_{\alpha}\psi_{\mu}~+~ghosts )
\label{1}
\end{equation}
Here
$\zeta =(\zeta_{1},~\zeta_{2})$ are world-sheet coordinates,
 $\hat{g}$ is a background world-sheet metric with curvature $\hat{R}$,
 $Q_{\mu}=(0,...,0,Q)$ is a background charge,
 $D_{\alpha}$ is covariant spin derivative;
$X_{\mu}$ and $\psi_{\mu}$ ($\mu=1,2, ... ,D$)
are $D$-dimensional matter fields $X_{\mu}=(X_i,\varphi)$, and
$\psi _{\mu}=(\psi _i,\psi)$, $(i=1,...,D-1)$,
where $X_i$ and $\psi_i$ are embedding bosonic and fermionic coordinates
and $\varphi$ and $\psi$ are the superLiouville modes. Signature
of the $D$-dimensional metric is chosen as $(++ \cdots +-)$.

The conformal energy-momentum tensor for such theory reads
\begin{equation}
T(z)=T^{X}(z)+T^{\psi}(z)+T^{bc}(z)+T^{\beta \gamma}(z),
\label{3}
\end{equation}
where $T^X$ and $T^{\psi}$ are energy-momentum tensors for $X_{\mu}$ and
$\psi_{\mu}$ fields
\begin{equation}
T^{X}=-\frac{1}{2}\partial X_{\mu}\partial X_{\mu}
-i\frac{1}{2}Q_{\mu}\partial ^{2}X_{\mu},
\label{4}
\end{equation}
\begin{equation}
T^{\psi}=\frac{1}{2}\psi_{\mu}\partial \psi_{\mu}
\label{5}
\end{equation}
and $T^{bc}$ and $T^{\beta \gamma}$ are energy-momentum tensors of
the spin (2,-1) $(bc)$-ghosts and the spin (3/2,-1/2) $(\beta \gamma)$-ghosts,
\begin{equation}
T^{bc}=-2b\partial c - \partial b c ,
\label{6}
\end{equation}
\begin{equation}
T^{\beta \gamma}=-\frac{3}{2}\beta \partial \gamma -\frac{1}{2}
\partial \beta \gamma .
\label{7}
\end{equation}
In all that follows we shall use the standard bosonization of the
superconformal
ghosts $\beta$ and $\gamma$ (see \cite{FMS})
\begin{equation}
\gamma=\eta e^{\phi},~~\beta=e^{-\phi}\partial \xi .
\label{8}
\end{equation}
The energy-momentum tensors for fields $\phi$ and $\eta \xi$ have the forms
\begin{equation}
T^{\phi}=-\frac{1}{2}\partial \phi \partial \phi -\partial ^{2} \phi
\label{9}
\end{equation}
\begin{equation}
T^{\eta \xi}=\partial \xi \eta.
\label{10}
\end{equation}
It is useful to write expressions for generators of the superconformal
transformations. We have
\begin{equation}
F^{X\psi}=-\frac{1}{2}(\psi _{\mu}\partial X_{\mu}
+iQ_{\mu}\partial \psi _{\mu})
\label{11}
\end{equation}
and
\begin{equation}
F^{bc,\beta \gamma}=-c\partial \beta  -\frac{3}{2} \partial
c\beta +\frac{1}{2} \gamma b
\label{12}
\end{equation}
for $x\psi$- and $bc\beta \gamma$- systems respectively.
 Superconformal invariance relates the background
charge $Q_{\mu}$ to the dimension $D$ of the target space as
\begin{equation}
D-2Q^2-10=0
\label{2}
\end{equation}

The BRST charge has the form
$$Q_{BRST}=\oint \frac{dz}{2\pi i}j(z),$$
\begin{equation}
j(z)=
c(T^{X}+T^{\psi}+\frac{1}{2}T^{bc}+T^{\xi \eta}+T^{\phi})(z)-
\eta e^{\phi}F^{X\psi}(z) + \frac{1}{4}b\partial \eta \eta
e^{2\phi}(z) .
\label{13}
\end{equation}
\subsection{Picture-changing operators}
Now let us write the expressions for the picture-changing operators for
the non-critical case. The first pair of picture-changing operators $X(z)$ and
$Y(z)$ can be constructed in standard manner as for the critical case
\cite{WSFT,FMS}. The operator $X(z)$ is represented as a BRST commutator in a
large
superconformal ghost algebra ($\xi$-algebra) and has the form
$$X(z)=\{Q_{BRST},\xi (z)\}=$$
\begin{equation}
=\frac{1}{2}e^{\phi}(\psi \cdot \partial X + iQ\cdot \partial \psi)(z)+
c\partial \xi (z)+
\frac{1}{4}b\partial \eta e^{2\phi}(z)
+\frac{1}{4}\partial(b \eta e^{2\phi})(z).
\label{14}
\end{equation}
The inverse operator $Y(z)$ has the same form as in the critical case
\begin{equation}
Y(z)=4c\partial \xi e^{-2\phi}(z)
\label{15}
\end{equation}
The double-steep picture-changing operators $W(z)$ and $Z(z)$   \cite{AMZ}
\begin{equation}
Z(z)W(z)=1,~Y(z)W(z)=X(z),~ X(z)Z(z)=Y(z),
\label{16}
\end{equation}
which suppose to be BRST invariant primary conformal fields
with zero conformal dimension
can also be found explicitly
$$W(z)=:X^{2}:(z)-$$
\begin{equation}
-\frac{19}{192}\{Q_{BRST}, \partial b\partial e^{2\phi}(z)\}
-\frac{5}{48}\{Q_{BRST}, \partial^{2} be^{2\phi}(z)\}
-\frac{1}{96}\{Q_{BRST}, b\partial ^{2}e^{2\phi}(z)\} ,
\label{17}
\end{equation}
\begin{equation}
Z(z)=-4e^{-2\phi}(z)-\frac{5}{8}c\partial \xi e^{-3\phi}(\psi \cdot
\partial X + iQ\cdot \partial \psi )(z).
\label{18}
\end{equation}
The properties (\ref{16}) can be easy  verified by OPE technique.
\section{String states.}
\subsection{ NS sector.}
In this section we consider string states in the bosonic sector of
the non-critical
NSR string theory. As in the critical case  such states can be constructed for
 different choices of superconformal ghost vacuum (different  pictures).

 Let us remind that the superconformal ghost vacuum $|\nu >$ in $\nu $-
 picture is connected with $SL_2$-invariant vacuum $|0>$ as
\begin{equation}
|\nu >=e^{\nu \phi (0)} |0> .
\label{19}
\end{equation}
There are two traditional pictures ($\nu =-1$ and $\nu =0$ )
in which NS states are
constructed.

For the critical theory the GSO condition is imposed on string states in both
NS
and R sectors. This condition excludes one half of states
and leads to the theory with
space-time supersymmetry. We must impose the GSO condition in the non-critical
case also in order to remove boson states with odd fermion number.
Vertex operators corresponding to the lowest level
 states in  $\nu =-1$ and $\nu =0$ pictures have the form
\begin{equation}
O_{-1}(z)=\zeta _{\mu}ce^{-\phi}\psi_{\mu}e^{ikX}(z),
\label{20}
\end{equation}
\begin{equation}
O_{0}(z)=\zeta _{\mu}[\frac{1}{2}c(\partial X_{\mu}
+ik\cdot \psi \psi _{\mu})-\frac {1}{4}\eta e^{\phi}\psi _{\mu}]e^{ikX}(z),
\label{21}
\end{equation}
where polarization vector $\zeta_{\mu}(k)$ satisfies the on-shell
conditions
\begin{equation}
(k^2+k\cdot Q)\zeta _{\mu}=0,
\label{22}
\end{equation}
\begin{equation}
(k+Q)\cdot \zeta =0.
\label{23}
\end{equation}
It is not difficult to verify that $O_{-1}(z)$
and $O_{0}(z)$ are BRST invariant primary conformal fields with zero
conformal dimension.

Let us note that the lowest level vertex operators (\ref{20})
and (\ref{21}) are nothing but the on-shell NS string field theory states
in the Siegel gauge.
In field theory the string states appear as a result of decomposition
of string field. In the NSR string field theory for the free case one can
work in one of these pictures.
For the free theory these pictures are equivalent and lead
to the same on-shell conditions on the component space-time fields \cite{UZ}.
However at the presence of an interaction the different choices of pictures
lead to different theories. It is known
that the NS string field theory in the $(-1)$-picture leads to
divergent tree level amplitudes
\cite{Wendt} and one can construct the Chern-Simons-like
superstring field theory only in the $(0)$-picture \cite{AMZ}.
In the first quantized approach it is more suitable to deal with vertex
operators in the $(-1)$-picture and put the necessary number of the
picture-changing operators on external states.

 Let us  consider decomposition of the NS string fields in both
  pictures. Hawing in mind the GSO condition and using the correspondence
  between states and vertex operators we write
\begin{equation}
|A_{\nu}>=A_{\nu}(0)|0>,
\label{24}
\end{equation}
where
$$ A_{-1}(z)=\int dk [A_{\mu}(k)ce^{-\phi}\psi _{\mu}+B(k)\partial cc
\partial \xi e^{-2\phi}+$$
\begin{equation}
+higher~~ level~~terms~~]e^{ikX}(z),
\label{25}
\end{equation}
$$ A_{0}(z)=\int dk [ \Phi (k) c + \frac{1}{2} A_{\mu}(k) c \partial
X_{\mu} - \frac{1}{4}B_{\mu}(k)\eta e^{\phi}\psi_{\mu} + \frac{1}{2}
F_{\mu \nu }(k)c\psi _{\mu} \psi _{\nu} + $$
\begin{equation}
+B(k)\partial c + F(k)c \partial \phi ~~+~~
higher~~ level~~ terms~~]e^{ikX}(z).
\label{26}
\end{equation}
It is important to stress that off-shell string fields (\ref{25})
and (\ref{26}) cannot be transformated in each other.
However the mass-shell equation $\{Q_{BRST},A_{\mu}(z)\}=0$
leads to the same conditions on the component fields for any picture:
$$(k^2+k\cdot Q)A_{\mu} - ik_{\mu}B=0,~~
i(k+Q)\cdot A+B=0,$$
\begin{equation}
A_{\mu}=B_{\mu},~~F_{\mu \nu}=\frac{1}{2}ik_{[\mu}A_{\nu ]},~~
F=0,~~\Phi =0 .
\label{27}
\end{equation}

The higher order excitation can be considered in the similar way,
however the expressions for vertex operators
are more complicate.

We are going make some comments about the specific case $D=2$.
It is instructive to recall the situation for the 2D bosonic string.
Naively one can expect that in $D=2$ case after gauge fixing survives only the
tachyonic state. However on mass-shell survive also the states with
discrete value of the momentum \cite{Pol,Lian,AZ,Pilch,IO}.
These states appear as a poles in the correlation functions \cite{GrKl,DiFr}
and in the scattering amplitudes \cite{Pol,AZ,Yang}
and are related with non-trivial cogomology of BRST charge
\cite{Lian,Pilch,IO}.
Let us sketch the appearance of these discreet states in
light-cone gauge.

As usual \cite{Thorn} one introduces
\begin{equation}
p^{\pm}=\frac{1}{\sqrt{2}}(p_{1}\pm p_{2}),~~~
\alpha ^{\pm}_{n}
=\frac{1}{\sqrt{2}}(
\alpha _{n1}\pm
\alpha _{n2}).
\label{a1}
\end{equation}
The mode expansion of the Virassoro operators
corresponding to the momentum-energy tensor $T(z)$
(\ref{3}) has the form
\begin{equation}
L_m=P^{+}(m)\alpha ^{-}_{m}+P^{-}(m)\alpha ^{+}_{m}
+\sum_{k\ne 0,m}\alpha ^{+}_{m-k}\alpha^{-}_{k}
\label{a2}
\end{equation}
for $m\ne 0$, where
\begin{equation}
P^{+}(m)=p^{+}+\frac{Q}{2\sqrt{2}}(m+1)
\label{n3}
\end{equation}
\begin{equation}
P^{-}(m)=p^{-}-\frac{Q}{2\sqrt{2}}(m+1)
\label{n4}
\end{equation}
 and
\begin{equation}
L_0=p^{+}p^{-}-
\frac{Q}{2\sqrt{2}}(p^{+}-p^{-})
+\sum_{k\ne 0}\alpha ^{+}_{-k}\alpha^{-}_{k}.
\label{a3}
\end{equation}
As in the usual case ($Q=0,~~ D=26$), to perform the quantization in the
light-cone gauge one has to fix $\alpha ^{+}_{m}=0$ for $m=\pm 1,2,...$,
then solve the  constraints
\begin{equation}
L_m=0,~~m\ne 0
\label{n5}
\end{equation}

For D=2 case  there is some subtlety in applying this procedure.
The solution of the constraints (\ref {n5}) becomes dependent of the
values of momentum and energy. If
$ P^{+}(m)$ are not equal to zero for all $m \ne 0$, then after fixing
$\alpha ^{+}=0$ the constraints (\ref{n5})
imply $\alpha ^{-}=0$ for all $m \ne 0$.
In this case one gets the single lowest scalar state which corresponds to
the so-called tachyon.

If there is such $m=m_0>0$
that $P^{+}(m_0)=0$ one can again fix $\alpha ^{+}_m=0$
and constraints (\ref{n5}) allow $\alpha ^{-}_{m_0}\ne 0$,
$~\alpha ^{-}_m=0$ for $m\ne m_0$.
To perform the quantization we have to assume that the
corresponding conjugated variables
$\alpha ^{+}_{-m_0}$ is also non-zero,
$[\alpha ^{\pm}_{m_0},\alpha ^{\mp}_{-m_0}]=m_0.
$
The constraints $L_m|phys.~states>=0$ for $m\ne m_0$ are
trivially satisfied. One can guarantee the condition
\begin{equation}
L_{m_0}|phys.~states>=
[P^{+}(m_0)\alpha^{-}_{m_0}+
P^{-}(m_0)\alpha^{+}_{m_0}
]|phys.~states>=0
\label{n7}
\end{equation}
 on states of the form
\begin{equation}
|phys.~states>=(\alpha^{+}_{-m_0})^k|p>
\label{n8}
\end{equation}
with $p$ such that the condition
\begin{equation}
P^{+}(m_0)=p^{+}+\frac{Q}{2\sqrt{2}}(m_0+1)=0
\label{n9}
\end{equation}
is satisfied. For bosonic case $Q=2\sqrt{2}$). The equation
\begin{equation}
(L_0-1)|phys.~states>=0
\label{nn}
\end{equation}
  with $L_0$ being (\ref{a3}) fixes the values of $p^{-}$ in
(\ref{n8})
\begin{equation}
p^{-}=k+1.
\label{nnn}
\end{equation}
So, the momentum and  energy in (\ref {n8}) take the following values:
\begin{equation}
p^{(I)}_1=
-\frac{1}{\sqrt{2}}(m_0-k),~~~p^{(I)}_2=
-\sqrt{2}
-\frac{1}{\sqrt{2}}(m_0+k).
\label{n16}
\end{equation}
We labelled this series of states by upper index $(I)$.

The case when $P^{-}(m_0)=0 $
for some $m_0$ can be considered analogously.
The condition (\ref {n7}) holds for all $m>0$ on the states
\begin{equation}
|phys.~states>=(\alpha^{-}_{-m_0})^k|p>
\label{n10}
\end{equation}
Equation (\ref {nn}) gives momenta of the
states (\ref{n10})
\begin{equation}
p^{(II)}_1=
\frac{1}{\sqrt{2}}(m_0-k),~~~p^{(II)}_2=
-\sqrt{2}
-\frac{1}{\sqrt{2}}(m_0+k).
\label{n17}
\end{equation}
Comparing (\ref{n16}) and (\ref{n17}) one can see that these series have
the same energy and opposite momentum. So, they describe the right and
left moving excitations with the same dispersion.
It is amusing to note that for the values of $p$ as in equations (\ref {n16})
$P^{-}(k)=0 $ holds.

Equation (\ref{n7}) is enough for   vanishing of matrix
elements of $L_{-m_0}$.
To see this, let us
remind the hermitian conjugation rules for operators $p$ and
$\alpha ^{\pm}_{m}$:
$p_1^{+}=p_1$, $p_2^{+}=-p_2-Q$,
$(\alpha ^{\pm}_{m})^{+}=
\alpha ^{\mp}_{-m}$.
With these rules we have $L_m^{+}=L_{-m}$
and states dual to
$(\alpha ^{\pm}_{-m_0})^k|p_1,p_2>$
have the form
$<p_1',p_2'|(\alpha ^{\mp}_{m_0})^k$, where
$p'=(p_1',p_2')$ denotes the eigenvalue for operator $p^{+}$:
$p_1'=p_1$, $p_2'=-p_2-Q$.
It is not difficult to see that on the dual states $<dual| L_{-m_0}=0$ holds.
Taking into account (\ref{n16}) and (\ref{n17})  for dual
states we have
\begin{equation}
p'^{(I)}_1=
-\frac{1}{\sqrt{2}}(m_0-k),~~~p'^{(I)}_2=
-\sqrt{2}
+\frac{1}{\sqrt{2}}(m_0+k),
\label{n18}
\end{equation}
 and
\begin{equation}
p'^{(II)}_1=
\frac{1}{\sqrt{2}}(m_0-k),~~~p'^{(II)}_2=
-\sqrt{2}
+\frac{1}{\sqrt{2}}(m_0+k)
\label{n19}
\end{equation}
respectively.
The similar situation occurs in the NS sector of the 2D fermionic string.
In this case due the superconformal invariance
one can assume that all components
$\alpha ^{+}_{m}$
and $\psi^{+}_s$
( or $\alpha ^{-}_{m}$
and $\psi^{-}_s$ )
are equal to zero.
In light-cone notations
 $\psi ^{\pm}_s=\frac{1}{\sqrt{2}}(\psi_{1s}\pm\psi_{2s})$
one can write the super Virassoro generators in the form
\begin{equation}
L_m=P^{+}(m)\alpha ^{-}_{m}+P^{-}(m)\alpha ^{+}_{m}
+\sum_{k\ne 0,m}\alpha ^{+}_{m-k}\alpha^{-}_{k}
+ \sum_{s}(s+\frac{1}{2})
\psi ^{+}_{m-s}\psi ^{-}_{s}
\label{n20}
\end{equation}
\begin{equation}
F_{s}=\frac{1}{2}[ K^{+}(s)\psi ^{-}_s+
K^{-}(s)\psi^{+}_s+\sum_{m\ne 0}
(\psi ^{+}_{s-m}\alpha^{-}_{m}+
\psi ^{-}_{s-m}\alpha^{+}_{m})]
\label{n21}
\end{equation}
Here $Q=2$, $P^{\pm}(m)$ are defined by (\ref{n3}), (\ref{n4}) and
\begin{equation}
K^{\pm}(s)=p^{\pm} \pm
\frac{Q}{\sqrt{2}}(s+\frac{1}{2}).
\label{n22}
\end{equation}
If both $P^{\pm}(m)$  and  $K^{\pm}(s)$
are  non-zero for all $m > 0$ and $s>0$   then in the light-cone gauge
$\alpha ^{+}_m=\psi ^{+}_s=0$
one has only trivial solution of constraint equations
$L_m=F_s=0$:
\begin{equation}
\alpha ^{-}_m=0,~~m\ne 0~;~~
\psi^{-}_s=0
\label{n23}
\end{equation}

The solution  (\ref{n23}) corresponds to a string state, which contains
only the lowest level scalar field.

However, if there is a such $m_0>0$ that
$P^{+}(m_0)=0$ then we have a non-trivial solution
$\alpha ^{+}_{m_0} \ne 0$.
For $P^{-}(m_0)=0$ then we have a non trivial solution
$\alpha ^{-}_{m_0} \ne 0$.
Analogously, if there is a such $s=s_0$ that
\begin{equation}
K^{\pm }(s_0)=p^{\pm}\pm \frac{Q}{\sqrt{2}}(s_0+\frac{1}{2})
=0
\label{n25}
\end{equation}
then we have
$\psi ^{\pm}_{s_0} \ne 0$.

The $m_0$- and $s_0$-modes produce
the following excitations:
\begin{equation}
(\alpha^{\pm}_{-m_0})^k|p>,
\label{n26}
\end{equation}
\begin{equation}
\psi^{\pm}_{-s_0}(\alpha^{\pm}_{-m_0})^k|p>,
\label{n27}
\end{equation}
The states (\ref{n27}) assume that the conditions (\ref{n3}) (or (\ref{n4}))
and
(\ref{n25}) hold simultaneously, hence here $s_0$ and $m_0$
should be related as
\begin{equation}
\frac{1}{2}(m_0+1)=s_0+\frac{1}{2}~~{\rm or}~~m_0=2s_0
\label{n28}
\end{equation}
Note that in the traditional
(-1)-picture the states (\ref{n26})
and lowest level scalar state must be exclude by the
GSO projection, since they describe the states with incorrect
statistic.
In (0)-picture
only the states of form (\ref{n26}) survive after the GSO projection.
Remind also that in our notations string  fields in the NS sector
assume to be fermionic.

The mass-shell equation
\begin{equation}
(L_0-\frac{1}{2})|phys.~states>=0
\label{n29}
\end{equation}
for physical states (\ref{n27})
gives the value of $p^{-}$ ($p^{+}$):
\begin{equation}
p^{\mp}=\pm \sqrt{2}(1+k).
\label{n30}
\end{equation}
So, one can see that in the NS sector after the GSO projection on the
$n$-mass level ($n=m_0k+s_0-\frac{1}{2}=$ $m_0(k+\frac{1}{2})-\frac{1}{2}$ )
only  states with discrete  momenta
\begin{equation}
p^{(I)}_1=\frac{1}{2}+\frac{1}{2}(2k-m_0),~~
p^{(I)}_2=-\frac{3}{2}-\frac{1}{2}(2k+m_0) ;
\label{n31}
\end{equation}
\begin{equation}
p^{(II)}_1=-\frac{1}{2}-\frac{1}{2}(2k-m_0),~~
p^{(II)}_2=-\frac{3}{2}-\frac{1}{2}(2k+m_0)
\label{n32}
\end{equation}
survive.

As for bosonic case we must add to the states (\ref{n27}) the dual states
\begin{equation}
<p_1',p_2'|
(\alpha^{\mp}_{m_0})^k
\psi^{\mp}_{s_0}
\label{n33}
\end{equation}
with
\begin{equation}
p'^{(I)}_1=\frac{1}{2}+\frac{1}{2}(2k-m_0),~~
p'^{(I)}_2=-\frac{3}{2}+\frac{1}{2}(2k+m_0);
\label{n34}
\end{equation}
\begin{equation}
p'^{(II)}_1=-\frac{1}{2}-\frac{1}{2}(2k-m_0),~~
p'^{(II)}_2=-\frac{3}{2}+\frac{1}{2}(2k+m_0).
\label{n35}
\end{equation}
\subsection{R sector}
Now we consider string states in
the fermionic sector of the non-critical NSR string.
As in case of the critical string such states are describes by spin field
vertex operators.
Such operators are expressed for
even $D$ in terms of anticommuting world-sheet fields
$\psi _{\mu}(z)$ when $\psi _{\mu}(z)$ is represented in bosonized form
(see  \cite{KLSW})
\begin{equation}
\psi _{2j-1}\pm i\psi _{2j}=e^{\pm \varphi _{j}}c_{j}.
\label{28}
\end{equation}
Here $\varphi _{j}(z) ~~(j=1, ...,\frac{D}{2})$ are free bosons
and $c_j$ are co-cycle operators,
which are, roughly speaking,  Jordan-Wigner factors
necessary to ensure that different fermions, when written in bosonized form,
anticommute.  The spin fields are represented as
\begin{equation}
S_{\alpha}(z)=e^{\lambda _{\alpha}\cdot \varphi}c_{\alpha},
\label{29}
\end{equation}
where
$\lambda_{\alpha}$
denotes an $\frac{D}{2}$-component spinor weight of form
$\lambda_{\alpha}=\frac{1}{2}(\pm,...,\pm)$.
There are $2^{\frac{D}{2}}$ possibilities, the correct
number for $SO(D)$-spinor.  One can decompose spin field $S_{\alpha}$ on two
chirality components,
which corresponds to two irreducible representation of $SO(D)$:
its positive (no dot on $\alpha$ ) and negative (dot on $\alpha$ )
chirality components.

Spin fields are operators with the conformal dimension $d(S_{\alpha})=
\frac{D}{16}$.
As in the critical $D=10$ case we can consider two chiral component of
$S_{\alpha}$ multiplied by superghost factor $e^{-\frac{\phi}{2}}$
$$S_{\alpha}e^{\pm \frac{\phi}{2}} ~~and~~
S_{\dot{\alpha}}e^{\pm \frac{\phi}{2}}. $$
One can consider $S_{\alpha}e^{-\frac{\phi}{2}}$  as a candidate on fermion
vertex. Conformal dimension of this operator is
$$ d=\frac{D}{16}+\frac{3}{8}=\frac{1}{8}Q^2+1.$$
To obtain the operator with conformal dimension one we need to multiply it
 by operator with $d=\frac{1}{8}Q^2$. This may be done by
 multiplying $S_{\alpha}e^{\frac{\phi}{2}}$ on $e^{iqX}$
 with $q$ satisfying
 $$ \frac{1}{2}q(q+Q)=-\frac{1}{8}Q^2,$$
\begin{equation}
(q+\frac{1}{2}Q)^2=0.
\label{33}
\end{equation}
The BRST invariant spin operator which corresponds to
state at lowest mass level has the form
\begin{equation}
O_{-\frac{1}{2}}(z)=\zeta ^{\alpha}ce^{-\frac{\phi}{2}}S_{\alpha}e^{iqX}(z)
\label{34}
\end{equation}
and corresponds to $\nu =-\frac{1}{2}$ picture.
The condition $\{Q_{BRST},O_{-\frac{1}{2}}\}=0$
implies the mass-shell condition
\begin{equation}
(q+\frac{1}{2}Q)_{\mu}\gamma _{\mu ~\beta}^{~\alpha}\zeta ^{\beta}=0
\label{35}
\end{equation}
on polarization spinor $\zeta ^{\alpha}$.
Eq. (\ref{35}) is the Dirac-like equation for
the lowest level R string state.

Acting on $O_{-\frac{1}{2}}$ by picture-changing operator $X(z)$
one gets the  BRST invariant lowest level spin operator in
$\frac{1}{2}$-picture.

Shifting $q\to q-\frac{1}{2}Q$ we get massless spinor equation. Hence,
at the lowest level we have the massless spinor field.  At higher levels there
are
 spin-vector fields. Let us examine these  higher  excitations for $D=2$ case.
 As in the case of the bosonic string and the NS sector
  naively one can expect that these spin-vector states
  can be removed by gauge transformation. However, these states can be
  removed  only if
  their momentum are not equal to some special values and once again we get a
  set of  discrete states.

Since we are interesting about spin-vector fields  with fermionic statistic
we assume that only some $\alpha$-operators survive after gauge fixing
and solving the constraints equation, i.e. the situation
is similar to the case of bosonic  string. Physical space is spanned by
\begin{equation}
(\alpha ^{+}_{-m_0})^k|\beta,p>,~~~
{\rm or}~~(\alpha ^{-}_{-m_0})^k|\beta,p>~
\label{b}
\end{equation}
(here $\beta$ in $|\beta ,p>$ is spinor index)
with the condition (\ref {n9}) on $p^{+}$,  or $P^{-}(m_0)=0$ on $p^{-}$.

The equation $L_0|phys.~states>=0$
implies
\begin{equation}
p^{+}p^{-}-\frac{Q}{2\sqrt{2}}(p^{+}-p^{-})-\frac{Q^2}{8}=-km_0
\label{b3}
\end{equation}
So, momenta of the states (\ref{b}) are given by
\begin{equation}
p^{(I)}_1=\frac{1}{2}(2k-m_0),~~
p^{(I)}_2=-1-\frac{1}{2}(2k+m_0);
\label{b5}
\end{equation}
\begin{equation}
p^{(II)}_1=-\frac{1}{2}(2k-m_0),~~
p^{(II)}_2=-1-\frac{1}{2}(2k+m_0).
\label{b6}
\end{equation}
Hence, in the R sector we have massless spinor fields
as  well as a set of  spin-vector field living only with fixed values of
momentum and energy.

Series of discrete states  dual to (\ref{b}) are
\begin{equation}
p'^{(I)}_1=\frac{1}{2}(2k-m_0),~~
p'^{(I)}_2=-1+\frac{1}{2}(2k+m_0).
\label{b7}
\end{equation}
\begin{equation}
p'^{(II)}_1=-\frac{1}{2}(2k-m_0),~~
p'^{(II)}_2=-1+\frac{1}{2}(2k+m_0).
\label{b8}
\end{equation}
\section{Supersymmetry}
We are going to examine the existence of  supersymmetry in the non-critical
NSR string theory.
Remind that in the critical theory supersymmetry takes place after the
GSO condition is imposed. Then at each mass level the number and the
masses
of states in the NS sector are equal to the number and the masses of states in
in the R sector. The situation in the non-critical theory is different.
Here equal level states in the NS and R sectors have different masses.
Consider for example the lowest level states in both sectors.
After redifinition of momentum
\begin{equation}
p=k+\frac{1}{2}Q
\label{42}
\end{equation}
the mass-shell conditions on polarization vector $\zeta _{\mu}$
and spinor $\zeta _{\alpha}$
 look as following
\begin{equation}
(p^2-\frac{1}{4}Q^2)\zeta _{\mu}=(p+\frac{1}{2}Q)\cdot \zeta =0,
\label{43}
\end{equation}
\begin{equation}
p^2\zeta _{\alpha}=(\hat{p}\zeta )_{\alpha}=0.
\label{44}
\end{equation}
Hence the lowest excitation in the NS sector is massive  while the
lowest excitation in the R sector is massless.
At $n$ level one has the following mass-shell condition
\begin{equation}
p^2+2n-\frac{1}{4}Q^2=0
\label{45}
\end{equation}
in the NS sector and
\begin{equation}
p^2+2n=0
\label{46}
\end{equation}
in the R sector.

 Hence, it is impossible to construct a linear local transformation
between the NS and R sectors which acts separately at each level.
So, an   analog of the critical NSR  supersymmetry transformation
  in the non-critical NSR theory
 does not exist.

This can be also clarified by following.
Let us consider  the
fermion vertex with positive chirality components which is obtained by the GSO
projection. As in the critical theory
such vertex can be chosen as a candidate for NS-R-symmetry
operator. The fermion vertex with zero momentum
has fractional conformal dimension and cannot be used as a
such operator. In order to obtain a spin operator with conformal dimension
one it is necessary to take a fermion vertex with momentum $q=-\frac{1}{2}Q$.
Then a candidate for NS-R-symmetry operator can be written as
\begin{equation}
\Sigma _{-\frac{1}{2}}=\oint \frac{dz}{2\pi i}\Theta ^{\alpha}
e^{-\frac{\phi}{2}}S_{\alpha}e^{iqX}(z).
\label{48}
\end{equation}
and
\begin{equation}
[Q_{BRST},\Sigma _{-\frac{1}{2}}]=0.
\label{48'}
\end{equation}
The fact that momentum of the fermionic vertex is non-zero
has rather non-trivial consequences.
First, this  means that NS-R transformations is given by a  non-local operator
that can be expected by comparing (\ref{43}) and
(\ref{44}).
Secondly, the transformation (\ref{48}) has meaning
on R vertices with the momenta being a subject of some restrictions.
In particular, for the lowest level R vertex
\begin{equation}
O_{-\frac{1}{2}}(z)=\zeta ^{\alpha}ce^{-\frac{\phi}{2}}
S_{\alpha}e^{ikX}(z)
\label{49}
\end{equation}
this restriction has the form
\begin{equation}
 k\cdot Q=-\frac{1}{2}Q^2.
\label{d}
\end{equation}
Indeed, acting by (\ref{48}) on the vertex (\ref{49})
we get
$$ [\Sigma _{-\frac{1}{2}},O_{-\frac{1}{2}}(z)]=$$
\begin{equation}
=\oint \frac{dw}{2\pi i}\Theta ^{\alpha}\zeta ^{\beta}(
ce^{-\phi}\hat{\psi} _{\alpha \beta}e^{i(k+q)X}(w)(w-z)^{-\frac{D}{8}+
\frac{1}{4}+kq }~+~less~singular~terms~).
\label{50}
\end{equation}
In order to get a meaningful expression similar to the NS vertex
we must have the first order pole in the first term
in (\ref{50}),
so, taking into account (\ref{2}), we must set
\begin{equation}
q\cdot k=\frac{1}{4}Q^2
\label{51}
\end{equation}
This equation fixes the Liouville  component $k_D$ of momentum $k$
in the R sector
$$ k_D=-\frac{1}{2}Q. $$
The expression (\ref{50}) with condition (\ref{51})  being  imposed,
becomes the NS vertex in (-1)-picture.
The momentum in (\ref{50}) after redefinition
 $p=k+q+\frac{1}{2}Q$
satisfies the equation
\begin{equation}
p^2=\frac{1}{4}Q^2.
\label{52}
\end{equation}

In D=2 case  this means that the transformation (\ref{48}) relates the NS
lowest  level excitation and the R lowest level excitation with fixed value of
the momentum. All the 0-level R states
(except the state with $p_2=-\frac{1}{2}Q$ ) have not
their own partners in the NS sector.

Acting  by the operator (\ref{48}) on the lowest level NS vertex in
the (0)-picture  (\ref{21}) we see that in order to  reproduce the structure
of the R vertex we  must  demand
\begin{equation}
k\cdot Q=0 ~~~{\rm or }~~~k_D=0
\label{dd}
\end{equation}
on states in the NS sector.
The operator (\ref{48}) together with their picture-changing  version
\begin{equation}
\Sigma _{\frac{1}{2}}
=\{Q_{BRST},\oint \frac{dz}{2\pi i}\Theta ^{\alpha}
\xi e^{-\frac{\phi}{2}}S_{\alpha}e^{iqX}(z)\}
\label{ddd}
\end{equation}
form the following algebra
\begin{equation}
[\Sigma ' _{\frac{1}{2}},
\Sigma _{-\frac{1}{2}} ]=
\Theta '^{\alpha}
\Theta ^{\beta}
\frac{1}{2}
\oint \frac{dz}{2\pi i}
(\partial {\hat X}-iQ\cdot \psi {\hat \psi })_{\alpha \beta}
e^{-iQX}.
\label{dddd}
\end{equation}
To summarize the above discussion,  we saw that although the non-critical
analogs of the SUSY transformations
form the closed algebra being  the natural generalization of the algebra
of supersymmetry, they
have sense on vertices
with fixed "energy".
\section{Non-critical NSR string field action}
Now we can write a NSR string field theory action for the non-critical
theory.
In fact, all of ingredients of the modified Witten version
of open superstring field theory,
including an associative $\ast $ product can be constructed for
non-critical case.
The form of action is the same as for the critical NSR
superstring \cite{AMZ}
$$ S_{NSR}=<(IA)(\infty)Z(i)Q_{BRST}A(0)>+
\frac{2}{3}<Z(z_0)(hA)(z_1)(hA)(z_2)(hA)(z_3)>+$$
\begin{equation}
+<(I\Psi)(\infty)Y(i)Q_{BPST}\Psi (0)>+
2<Y(z_0)(h\Psi )(z_1)(hA )(z_2)(h\Psi )(z_3)>.
\label{53}
\end{equation}
Here $I$ is the map
$z \to z'=-\frac{1}{z}$ from the inside of unit disk to the
outside and $h$ is the map from interaction three-string Witten's
configuration to upper half-plane. The NS string field $A$
and the R string field $\Psi$ are built under vacuum with picture $\nu =0$
and $\nu =-\frac{1}{2}$ respectively.

The action (\ref{53}) is gauge invariant.
The form of gauge transformation is the same
 as for the critical NSR superstring field theory.
 However in our case the theory is not supersymmetric in the usual sense.
 As it was discussed in section 4 the absence of supersymmetry
 takes place due to the fact that equal level excitations
 in the NS and R sectors have different masses.

Let us make some comments about an action for local fields which can be
extracted from the action (\ref{53}).  The free action for the lowest
right (left)
excitation $\psi _{\pm}(x)$ in the 2D
R sector which comes from the first term of the second line
in (\ref{53}) has the form
\begin{equation}
S_{0, \psi }= \int d^{D}x e^{-2ix_D}
\psi _{\pm}(x)( i\partial _1 \pm i\partial _2 \mp 1)\psi _{\pm}(x).
\label{c}
\end{equation}
 As to the free action for lowest lying fields in the NS sector there the
situation is more complicate since there are auxiliary fields
(compare with the critical case \cite{UZ}).

  The string field action (\ref{53}) contains  an interaction between
massless spinor and higher level fields. Since the action   (\ref{53})
does not contain the terms describing $\psi ^{4}$ interaction, $\psi ^{4}$
term comes from the interaction between massless spinor and others
fields. Contributions to $\psi ^{4}$  term  can been calculated using
off-shell conformal methods or level truncation approximation  \cite {Kost}.
For example the $g^{2}$-order contributions can be written schematically as
\begin{equation}
S_{int,\psi}~=~g^2
\int Y \Psi \ast \Psi
\ast \frac{b_0}{L}X(\Psi \ast \Psi)
-g^2\int Y \Psi \ast \Psi
\ast \frac{b_0}{L}W\frac{b_0}{L}YQ(\Psi \ast \Psi) .
\label{cc}
\end{equation}

\section{Tree amplitudes}

We shall employ the formalism of string field  theory
to get string amplitudes.
The Liouville degree of freedom make the Witten vertex
BRST invariant, and the $\ast $ product associative.
These features will then guarantee duality and tree-level
unitarity (factorisation).

 Let us note that principles of calculation of tree amplitudes are
 the same as in the critical
 superstring field theory. Remind the main steps of this calculations
 \cite{GM,BS}.
 To invert the kinetic operator we must fix the gauge. A convenient
 choice is the Siegel gauge $b_0A=b_0\Psi =0$ where $b_0$ is
 the zero mode of $b(z)$.
 The propagators in this gauge are
$$\Delta _{NS}=\frac{b_0}{L}Q_{BRST}W(i)\frac{b_0}{L}$$
in the NS sector and
$$\Delta _{R}=\frac{b_0}{L}Q_{BRST}X(i)\frac{b_0}{L}$$
in the R sector. With each Feynman graph a string configuration is associated.
External string are semi-infinite rectangular strips of width $\pi$.
The effect of $\frac{1}{L}=$ $\int _{0}^{\infty}d\tau e^{-\tau L}$
in propagators is to introduce a world-sheet strip of width $\pi$ and length
$\tau$. The interaction glues the strip end together in pairwither manner.

It may be shown that contribution to a particular Feynman graph is
\begin{equation}
A_{N}=(\prod_{i=1}^{N-3}\int d\tau _{i})
<\prod _{r=1}^{N}O^{(r)}(w_r)
\prod _{i=1}^{N-3}\oint dw_i'b(w_i')
\prod _{j}{\cal Z}(w_j'')>_{R_{\tau}},
\label{54}
\end{equation}
Where ${\cal Z}_{j}(w_{j}'')$ are appropriate picture-changing
operator insertions.
The correlation function (\ref{54})
is considered on the string configuration $R_{\tau}$
described above.

Analysis of field theory tree level amplitudes (\ref{54})
for the non-critical case is completely the same as for
the critical theory \cite{AMZ}.
The result is that all tree-level amplitudes are finite and
can be presented in the first quantized language.
\subsection{Four-fermion amplitude}
The amplitude for four lowest level  states in the R sector  can be
represented as
\begin{equation}
A_{4F}=\int _{z_4}^{z_2}dz_3<
O^{(1)}_{-\frac{1}{2}}(z_1)
O^{(2)}_{-\frac{1}{2}}(z_2)
V^{(3)}_{-\frac{1}{2}}(z_3)
O^{(4)}_{-\frac{1}{2}}(z_4)>
\label{55}
\end{equation}
Here $O_{-\frac{1}{2}}(z)$
are vertex operators in $(-\frac{1}{2})$-picture
which have the following form
\begin{equation}
O^{(r)}_{-\frac{1}{2}}(z_r)=cV^{(r)}_{-\frac{1}{2}}(z_r)=
\zeta ^{(r)\alpha}ce^{-\frac{\phi}{2}}
S_{\alpha}e^{ik^{(r)}X}(z_r).
\label{56}
\end{equation}
The polarization spinors $\zeta ^{(r)\alpha}$
satisfy the mass-shell condition
\begin{equation}
(k^{(r)}+\frac{1}{2}Q)^2\zeta ^{(r)\alpha}=(k^{(r)}
+\frac{1}{2}Q)^{\alpha}_{~\beta}
\zeta ^{(r)\beta}=0
\label{57}
\end{equation}
The points $z_r$ are arbitrary and lie on real axis so that $z_1
>z_2>z_3>z_4$.
Usual choice is $z_1=\infty,~z_2=1,~z_4=0$. It is not difficult
to verify that the correlation function (\ref{55}) may be represented in the
form
\begin{equation}
A_{4F}=\int _{0}^{1}dx x^{k^{(3)}k^{(4)}-\frac{1}{4}Q^2-1}
(1-x)^{k^{(2)}k^{(3)}-\frac{1}{4}Q^2-1}K(\zeta ^{(r)},k^{(r)},x),
\label{58}
\end{equation}
where function $K$ is depended on polarization spinors $\zeta ^{(r)}$,
external momentums $k^{(r)}$ and is linear under $x$.

One can introduce $s$-  and $t$- channel variables
\begin{equation}
s=(k^{(1)}+k^{(2)}+\frac{1}{2}Q)^2,~~~t=(k^{(2)}+k^{(3)}+\frac{1}{2}Q)^2.
\label{59}
\end{equation}
On mass-shell s and t variables  can be written  as
\begin{equation}
s=2k^{(1)}k^{(2)}-\frac{1}{4}Q^2,~~~t=2k^{(2)}k^{(3)}-\frac{1}{4}Q^2
\label{60}
\end{equation}
Taking into account the relation $ k^{(1)}k^{(2)}=k^{(3)}k^{(4)} $
one can represent
the amplitude (\ref{58}) in the form similar to the well known form
of fermionic
4-point amplitude
\begin{equation}
A_{4F}=\int _{0}^{1}dx x^{\frac{1}{2}s-\frac{1}{8}Q^2-1}
(1-x)^{\frac{1}{2}t-\frac{1}{8}Q^2-1}K(\zeta ^{(r)},k^{(r)},x)
\label{61}
\end{equation}
To study a pole structure of $A_{4F}$ we set simply $K=1$. Then
\begin{equation}
A_{4F}\sim \frac{\Gamma (\frac{1}{2}s-\frac{1}{8}Q^2)
\Gamma (\frac{1}{2}t-
\frac{1}{8}Q^2)}{\Gamma (\frac{1}{2}s+\frac{1}{2}t-\frac{1}{4}Q^2)}
\label{62}
\end{equation}
The four-fermion amplitude has  $s$- and $t$-
 poles  which correspond to the  excitations in the NS sector:
\begin{equation}
s=\frac{1}{4}Q^2-2n,~~t= \frac{1}{4}Q^2-2n,~~(n=0,1,2,...~)
\label{63}
\end{equation}

For $D=2$ case due
to the special kinematical relations this amplitude can be presented as a
function of the individual external momentum. Indeed,
the mass-shell condition for external states,
\begin {equation} 
\label {64}
k_{2}^{(r)}=-1+e_rk^{(r)},~~~e_{r}=\pm 1
\end   {equation} 
and the momentum-energy conservation low
\begin {equation} 
 \sum _{r=1}^{4} k_1^{(r)}=0,~~~\sum _{r=1}^{4}k_2^{(r)}=-Q
\label{66}
\end   {equation} 
in the particular case $e_{i}=1,~i=1,2,3,~e_{4}=-1,$
(three right-moving and one   left-moving fermions) give
$$k_1^{(4)}=-1,~~~k_1^{(1)}+k_1^{(2)}+k_1^{(3)}=1.$$
and
$$s=-1-2(k_1^{(1)}+k_1^{(2)}),~~t=-1-2(k_1^{(2)}+k_1^{(3)}),$$
and therefore
\begin {equation} 
\label {67}
 A^{4F}=\frac{\Gamma (-2+k_1^{(3)})
\Gamma (-2+k_1^{(1)})}{\Gamma (-3-k_1^{(2)})}
\end   {equation} 

This amplitude has the s-poles
which correspond to the state with fixed momentum and energy
\begin {equation} 
\label {68}
k_1=n ,~~k_2 =-1\pm (n+1),~~n=0,1,...
\end   {equation} 
Comparing (\ref{68}) with formulas (\ref{n32}) and (\ref{n35})
for $k=0$ one can see that these poles correspond to a part of
the set of discrete states in the NS sector:
\begin {equation} 
\label {69}
\psi ^{-}_{-m+\frac{1}{2}}|k_1,k_2 >~~{\rm  and} ~~
\psi ^{+}_{-m+\frac{1}{2}}|k_1,-k_2 -Q>.
\end   {equation}

Note that all scattering amplitudes for the fermions
of the same chirality (i.e. only right-moving or left-moving) are forbidden by
the momentum-energy conservation low.
Therefore after the GSO projection we left with trivial $N$-point  fermionic
amplitudes.
\subsection{Two-boson-two-fermion amplitude}
To calculate the scattering amplitude of two lowest level
bosons and two lowest level fermions we consider
the following four vertex correlation function
\begin{equation}
A_{2B2F}=\int _{z_4}^{z_{2}}dz_{3}<
O^{(1)}_{0}(z_1)
O^{(2)}_{-1}(z_2)
V^{(3)}_{-\frac{1}{2}}(z_3)
O^{(4)}_{-\frac{1}{2}}(z_4)>
\label{102}
\end{equation}
where
\begin{equation}
O^{(2)}_{-1}(z_2)=\zeta ^{(2)}_{\mu}ce^{-\phi}\psi_{\mu}e^{ik^{(2)}X}(z_2),
\label{103}
\end{equation}
\begin{equation}
O^{(1)}_{0}(z_1)=\zeta ^{(1)}_{\mu}[\frac{1}{2}c(\partial X_{\mu}
+ik^{(1)}\cdot \psi \psi _{\mu})-\frac {1}{4}\eta e^{\phi}\psi _{\mu}]
e^{ik^{(1)}X}(z_1),
\label{104}
\end{equation}
 $O^{(4)}_{-\frac{1}{2}}(z_4)=cV^{(3)}_{-\frac{1}{2}}(z_3) $
is given by (\ref{46}).
The polarization spinors $\zeta ^{(j)\alpha}$, $j=3,4$
and the polarization vectors $\zeta ^{(i)}_{\mu}$, $i=1,2$
satisfy to the mass-shell conditions (\ref{57}) and
\begin{equation}
((k^{(i)}+\frac{1}{2}Q)^2-\frac{1}{4}Q^2)\zeta ^{(i)}_{\mu}=(k^{(i)}
+Q)_{\mu}\zeta ^{(i)}_{\mu}=0
\label{105}
\end{equation}
respectively.

After the OPE calculations one gets
\begin{equation}
A_{2B2F}=\int _{0}^{1}dx x^{k^{(3)}k^{(4)}-\frac{1}{4}Q^2-1}
(1-x)^{k^{(2)}k^{(3)}-1}K(\zeta ^{(r)},k^{(r)},x)
\label{106}
\end{equation}
The on-shell amplitude (\ref{106}) being written in the $s$-  and $t$- channel
variables
\begin{equation}
s=2k^{(1)}k^{(2)}+\frac{1}{4}Q^2,~~~t=2k^{(2)}k^{(3)}.
\label{107}
\end{equation}
due to the kinematical relation $ k^{(1)}k^{(2)}=k^{(3)}k^{(4)}
-\frac{1}{4}Q^2$, has the form
$$A_{2B2F}=\int _{0}^{1}dx x^{\frac{1}{2}s-\frac{1}{8}Q^2-1}
(1-x)^{\frac{1}{2}t-1}K(\zeta ^{(r)},k^{(r)},x) ~\sim$$
\begin{equation}
\sim~ \frac{\Gamma (\frac{1}{2}s-\frac{1}{8}Q^2)
\Gamma (\frac{1}{2}t)}{\Gamma (\frac{1}{2}s+\frac{1}{2}t-\frac{1}{8}Q^2)}
\label{70}
\end{equation}
This amplitude  has  $s$- and $t$ poles               at
\begin{equation}
s=\frac{1}{4}Q^2-2n, ~~t=-2n,~~(n=0,1,2,...~)
\label{71}
\end{equation}
For the D=2 case the amplitude (\ref{71})
does not contains poles connected with discrete states
(compare with 2-vector-2-tachyon scattering  amplitude for the
bosonic case \cite{AZ}).
\subsection{Four boson amplitude}
Four boson scattering amplitude
\begin{equation}
A_{4B}=\int _{z_1}^{z_2}dz_3<
O^{(1)}_{0}(z_1)
O^{(2)}_{-1}(z_2)
V^{(3)}_{-1}(z_3)
O^{(4)}_{0}(z_4)>
\label{72}
\end{equation}
can be present in the form
\begin{equation}
A_{4B}=\int _{0}^{1}dx x^{k^{(3)}k^{(4)}-\frac{1}{4}Q^2-1}
(1-x)^{k^{(2)}k^{(3)}-\frac{1}{4}Q^2-1}K(\zeta ^{(r)},k^{(r)},x)
\label{73}
\end{equation}
In $s$- and $t$- variables this amplitude can be rewritten as
$$A_{4B}=\int _{0}^{1}dx x^{\frac{1}{2}s-\frac{1}{8}Q^2-1}
(1-x)^{\frac{1}{2}t-\frac{1}{8}Q^2-1}K(\zeta ^{(r)},k^{(r)},x).$$
Dropping the kinematical factor $K$ one get
\begin{equation}
A_{4B}~\sim~ \frac{\Gamma (\frac{1}{2}s-\frac{1}{8}Q^2)
\Gamma (\frac{1}{2}t-\frac{1}{8}Q^2)}
{\Gamma (\frac{1}{2}s+\frac{1}{2}t-\frac{1}{4}Q^2)}
\label{74}
\end{equation}
This amplitude has  $s$- and $t$- poles at
\begin{equation}
s=\frac{1}{4}Q^2-2n, ~~
t=\frac{1}{4}Q^2-2n, ~~~(n=0,1,2,...~)
\label{75}
\end{equation}

\newpage
\appendix
\section{APPENDIX. OPE for  spin fields for non-critical dimension}
The operator product expansion of two $S_{\alpha}$ can be written as
$$ S_{\alpha}(z) S_{\beta}(w)=\frac{1}{(z-w)^{\frac{D}{8}}}C_{\alpha \beta}
+$$
\begin{equation}
+\frac{1}{(z-w)^{\frac{D}{8}-\frac{1}{2}}}\gamma _{\mu \alpha \beta}
\psi _{\mu}(w)+\frac{1}{(z-w)^{\frac{D}{8}-1}}\frac{1}{2}
\gamma _{\mu \alpha}^{~~\gamma}\gamma _{\nu \gamma \beta}
\psi_{\mu}\psi _{\nu}(w) ~+~...~.
\label{30}
\end{equation}
The coefficients $c_{\alpha \beta}$ and $\gamma _{\alpha \beta}$
are defined by (\ref{30}).
The indices $\alpha$ and $\dot{\alpha}$
correspond to $\lambda _{\alpha}$ with even and odd number of
minuses respectively.
Situation will be different in dependences of would be $\frac{D}{2}$
even or not.

If $\frac{D}{2}$ is even, the leading singularity in (\ref{30})
will appear when indices $\alpha$ and $\beta$ correspond to same chirality.
Then we have
 $$ S_{\alpha}(z) S_{\beta}(w)=\frac{1}{(z-w)^{\frac{D}{8}}}C_{\alpha \beta}
 ~~+~~O((z-w)^{-\frac{D}{8}+1})~,$$
\begin{equation}
S_{\alpha}(z)S_{\dot{\beta}}(w)=\frac{1}{(z-w)^{\frac{D}{8}
-\frac{1}{2}}}\gamma _{\mu \alpha \beta{\beta}}
\psi _{\mu}(w)~~+~~O((z-w)^{-\frac{D}{8}+\frac{3}{2}})~.
\label{31}
\end{equation}
In this case the $2^{\frac{D}{2}-1}$ operators $S_{\alpha}$
and $2^{\frac{D}{2}-1}$ operators $S_{\dot{\alpha}}$ are transformed
are the two different spinor representation of $SO(D)$.

When $\frac{D}{2}$ is odd, we have
 $$ S_{\alpha}(z) S_{\dot{\beta}}(w)
 =\frac{1}{(z-w)^{\frac{D}{8}}}C_{\alpha \dot{\beta}}
 ~~+~~O((z-w)^{-\frac{D}{8}+1})~,$$
\begin{equation}
S_{\alpha}(z)S_{\beta}(w)=\frac{1}{(z-w)^{\frac{D}{8}-\frac{1}{2}}}\gamma _{\mu
\alpha \beta}
\psi _{\mu}(w)~~+~~O((z-w)^{\frac{D}{8}+\frac{3}{2}})~
\label{32}
\end{equation}
and $S_{\alpha}$ and $S_{\dot{\alpha}}$
are transformated as isomorphic representations.

Formulas (\ref{31}) and (\ref{32}) define the charge-conjugation
matrix $C$ , which reduces to the spinor metric.
For even $\frac{D}{2}$ $C$ split into two spinor metrics $C_{\alpha \beta}$
and $C_{\dot{\alpha} \dot{\beta}}$.
For odd $\frac{D}{2}$ we have $C_{\alpha \dot{\beta}}=-C_{\dot{\beta} \alpha}$.
We remind that in chiral basis the gamma matrices have one upper and one
lower index, one dotted and one undotted. For any $\frac{D}{2}$
indices may be raised and lowered with the appropriate spinor metric.

\end{document}